\documentclass[reprint,superscriptaddress,amsmath,amssymb,aps, prl,colorlinks]{revtex4-2}
\usepackage{graphicx}
\usepackage{dcolumn}
\usepackage{bm}
\usepackage[usenames,dvipsnames]{color}
\usepackage{natbib}
\usepackage{amsmath}
\usepackage{amssymb}
\usepackage{booktabs}
\usepackage{textcomp}
\usepackage[urlcolor=NavyBlue,citecolor=NavyBlue,linkcolor=NavyBlue]{hyperref}

\newcommand{\mean}[1]{\left < #1 \right >}

\renewcommand{\vec}[1]{\mathbf{ #1 }}

\makeatletter

\begin{document}

\title{Intermittent Run Motility of Bacteria in Gels \\ Exhibits Power-Law Distributed Dwell Times}

\author{Agniva Datta}
\affiliation{Institute of Physics and Astronomy, University of Potsdam, D-14476 Potsdam, Germany}

\author{S{\"o}nke Beier}
\affiliation{Institute of Physics and Astronomy, University of Potsdam, D-14476 Potsdam, Germany}

\author{Veronika Pfeifer}
\affiliation{Institute of Physics and Astronomy, University of Potsdam, D-14476 Potsdam, Germany}

\author{Robert Gro{\ss}mann}
\affiliation{Institute of Physics and Astronomy, University of Potsdam, D-14476 Potsdam, Germany}

\author{Carsten Beta}
\email[Correspondence should be addressed to ]{carsten.beta@uni-potsdam.de}
\affiliation{Institute of Physics and Astronomy, University of Potsdam, D-14476 Potsdam, Germany}
\affiliation{Nano Life Science Institute (WPI-NanoLSI), Kanazawa University, Kakuma-machi, Kanazawa 920-1192, Japan}

\begin{abstract}
While bacterial swimming has been well characterized in uniform liquid environments, only little is known about how bacteria propagate through complex environments, such as gel-like matrices or porous media that are typically encountered in tissue or soil. 
Here, we study swimming motility of the soil bacterium {\it Pseudomonas putida}~({\it P.~putida}) in polysaccharide matrices formed by different concentrations of agar. 
{\it P.~putida} cells display intermittent run-motility in the gel, where run times are exponentially distributed and intermittently occurring dwell times follow a waiting-time distribution with a power-law decay. 
An analysis of the turn angle distribution suggests that both, flagella mediated turning as well as mechanical trapping in the agar matrix play a role in the overall swimming pattern. 
Based on the experimentally observed motility pattern and measured waiting-time distributions, we propose a minimal active particle model which correctly describes the observed time dependence of the mean square displacement of the bacterial swimmers.
\end{abstract}

\date{\today}
\maketitle

Bacterial swimming is one of the most common forms of motility at the cellular level~\cite{wadhwa2022bacterial}.
For many decades, swimming in uniform liquid environments has been studied, mostly focusing on the classical example of {\it E.~coli} that moves in a sequence of straight runs, interrupted by motor-induced abrupt tumbling events~\cite{berg2004coli}.
However, in many natural habitats, such as mucus, soil, or plant and animal tissues, bacterial swimmers face heterogeneities and strong confinement.
Compared to free swimming in a uniform liquid, much less is known about how individual cells navigate such complex environments.

To address this question, bacterial swimmers were studied in well-defined microfluidic geometries and other artificial compartments~\cite{mannik2009bacterial,tokarova2021patterns,du2021bacterial}, taking also the impact of fluid flow into account~\cite{conrad2018confined}.
Also, disordered porous media were considered, in which {\it E.~coli} cells move in directed hops that are interrupted by transient mechanical trapping events in narrow spaces or cavities of the porous material~\cite{bhattacharjee2019bacterial,bhattacharjee2019confinement}. In many natural habitats, however, such as the human gut or the plant rhizosphere, bacteria move through layers of mucus and other gelatinous matrices. To date, a close analysis is missing that shows how bacterial swimming patterns change in such gel-like surroundings. It is a key prerequisite to understand how pathogens invade host cells and tissue~\cite{balzan2007bacterial,ribet2015bacterial,chaban2015flagellum,magold2022pathogenic} or how rhizobia colonize plant roots in the soil~\cite{turnbull2001role,tecon2017biophysical,aroney2021rhizobial}.

Studies of bacterial swimming in bulk liquid have shown that, besides the classical run-and-tumble paradigm of {\it E.~coli}, a variety of different swimming modes and turning maneuvers can be observed in other species.
During runs of peritrichously flagellated {\it E.~coli}, cells always move in push mode, with the flagellar bundle propelling the cell body from behind.
In contrast, polarly flagellated species may also switch to pull mode or perform screw thread motility with their flagellar helix wrapped around the cell body~\cite{thormann2022wrapped}.
Sharp directional reversals, flagellar flicking and wrapping are turning maneuvers observed along with these different swimming modes~\cite{grognot2021more}.
How they affect motility in complex environments is largely unexplored.

Here, we study motility of the soil bacterium {\it Pseudomonas putida} ({\it P.~putida}) based on single cell trajectory recordings in a gel-like polysaccharide matrix formed by semisolid agar.
{\it P.~putida} propels itself with a tuft of flagella polarly attached to one end of the rod-shaped cell body~\cite{harwood1989flagellation}.
It can swim in push, pull, or wrapped mode, interrupted by directional reversals or stop events~\cite{Theves2013,Hintsche2017}.
The wrapped mode was observed to play a key role in chemotaxis~\cite{alirezaeizanjani2020chemotaxis}, it is more likely to occur under conditions of increased viscosity and mechanical confinement, and its occurrence can be tuned by introducing knockout mutations in the torque generating MotAB and MotCD stator units of the flagellar motor~\cite{Pfeifer2022}.

We recorded the motility of {\it P.~putida} KT2440 wild type cells, as well as $\Delta$\textit{motAB} and $\Delta$\textit{motCD} mutant cells in semisolid agar with phase-contrast microscopy at a frame rate of 20 frames per second.
Recordings of a total duration of $60 \, \mbox{s}$ were segmented to identify the positions of cells in each time frame.
Cell trajectories were then extracted from the time lapse image sequences by linking cell positions in successive frames using a next neighbor particle tracking algorithm.
In addition, we performed high-resolution dual color fluorescence microscopy at a rate of 100 frames per second to identify the different swimming modes.
Here, the flagella were labeled with Alexa 488 C$_5$ maleimide and the cell body was stained with a red membrane intercalating dye.
For details of cell culture, microscopy recordings, and image analysis, see Supplemental Material~\cite{noteSI}.

\paragraph{Bacterial swimmers display intermittent run motility in agar gel.}

\begin{figure}[t!]
    \includegraphics[width=\columnwidth]{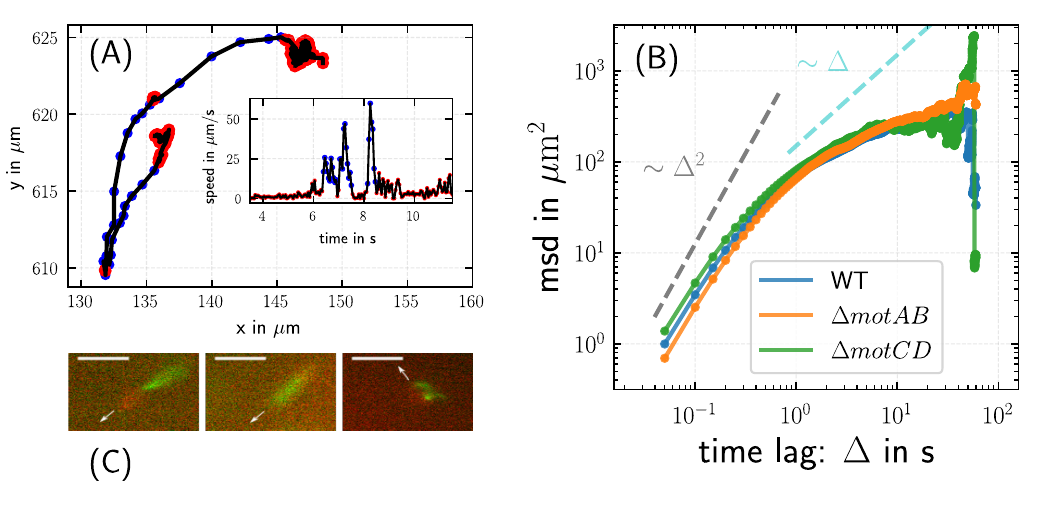}
    \caption{Trajectories reveal intermittent run-motility of bacteria in agar gel.~(A)~Trajectory of a bacterium in $0.25 \,\%$ agar, with the corresponding speed as a function of time as an inset. Runs are indicated in blue. (B)~MSD of \textit{P.~putida} wild type, $\Delta$\textit{motAB}, and $\Delta$\textit{motCD} in $0.25\,\%$ agar. (C)~The three swimming modes push, pull, and wrapped in semi-solid agar shown for fluorescently labeled cells. The scale-bar corresponds to $5 \, \mbox{\textmu m}.$ }
    \label{fig:trajectory}
\end{figure}

Visual inspection of trajectories revealed that they are typically composed of straight runs, interrupted by phases during which the bacterium hovers around a fixed position before performing the next run in a different direction, see Fig.~\ref{fig:trajectory}A for an example trajectory.
Given the extended duration of many of these events in comparison to the short turn maneuvers~(stops and reversals) observed in open liquid~\cite{Theves2013,alirezaeizanjani2020chemotaxis}, we assume that these events are associated with mechanical trapping in the gel matrix, where cells may only perform a confined wiggling motion until they free themselves again.
In order to separate runs from immobile episodes, we applied k-means clustering~\cite{bishop2006} on the speed of the bacterium and the change of direction of motion~(see Supplemental Material~(SM) for details~\cite{noteSI}). 
Furthermore, our fluorescence microscopy recordings demonstrated that all three swimming modes~--~push, pull, and wrapped~--~that are known from motility in bulk liquid are also present in the porous environment, see Fig.~\ref{fig:trajectory}C.

\paragraph{Mean square displacement shows robust crossover from ballistic to sublinear scaling.}

The mean square displacements~(MSD)~$m_2 \! \left ( t_a; \Delta \right ) = \langle \left |  \mathbf{r}(t_a+\Delta) - \mathbf{r}(t_a) \right |^2\rangle$ of \textit{P.~putida} wild type, $\Delta$\textit{motAB}, and $\Delta$\textit{motCD} mutant cells are displayed in Fig.~\ref{fig:trajectory}B.
For all three cell lines, the time evolution of the MSD shows a similar characteristic, namely a crossover around $\Delta \approx 0.5 \, \mbox{s}$ from ballistic motion with $m_2 \sim \Delta^{\!\!\:2}$ at short times to a sublinear regime at long times. 
As the three swimming modes occur with different frequencies in wild type and mutant cells~\cite{Pfeifer2022}, the similar shape of the MSD curves indicates that bacterial spreading in the agar matrix does not depend on the frequency of occurrence of different swimming modes but is mainly determined by the structure of the matrix itself.

\paragraph{Transient trapping in agar vs.~flagella mediated bulk motility.}

\begin{figure}[b]
    \includegraphics[width=\columnwidth]{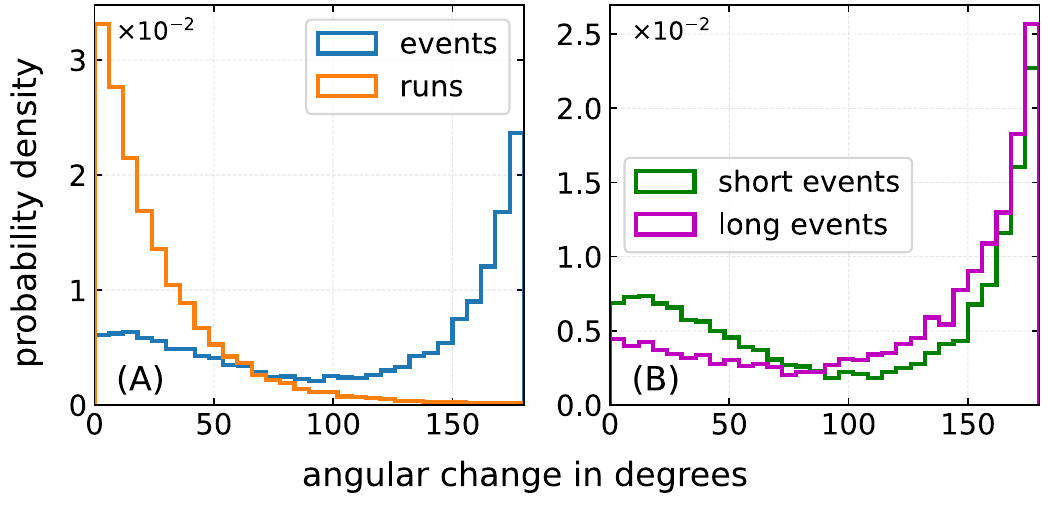}
    \caption{(A) Change of the direction of motion during runs~(orange) and events~(blue). (B)~Change of the direction of motion during short~($\leq 1 \, \mbox{s}$, shown in green) and long~($>1 \, \mbox{s}$, purple line) turn events for {\it P.~putida} wild type in $0.25 \, \%$ agar.}  
    \label{fig:angular_change}
\end{figure}

Measuring the change in propagation direction during run and immobile phases, we found that runs, as expected, exhibit a high directional persistence, i.e.~the angular change during a run remains small and peaks around $0^{\circ}$, see Fig.~\ref{fig:angular_change}A. 
On the other hand, large directional changes are expected when trapping is caused by dead ends and cavities. 
In these cases, cells have to leave the trap via the same way through which they entered, resulting in a directional reversal, i.e.~an angular change of $180^{\circ}$. 
A pronounced peak at $180^{\circ}$ in the distribution of directional changes during turn events confirms this expectation, see Fig.~\ref{fig:angular_change}A.

Besides a peak at $180^{\circ}$, the turn-angle distribution displays a second peak centered around $0^{\circ}$~(Fig.~\ref{fig:angular_change}A). 
This is reminiscent of the bimodal turn angle histogram that was observed for {\it P.~putida} when swimming freely in a uniform liquid environment~\cite{Theves2013}.
It suggests that the events identified by our distinction criteria may contain not only trapping in the gel matrix, but also actively triggered stop and reversal events that occur as part of the intrinsic swimming pattern of {\it P.~putida}.
While cells may remain mechanically trapped for several seconds, however, actively triggered stops and turns are typically much shorter than $1 \, \mbox{s}$~\cite{Theves2013,alirezaeizanjani2020chemotaxis}. 
To achieve an approximate distinction of mechanical trapping and active turn events, we split all events detected by k-means clustering into two types, those with a duration longer and those shorter than $1 \, \mbox{s}$, respectively. 
In Fig.~\ref{fig:angular_change}B, the corresponding distributions of directional changes are displayed for long and short events.
For short events, we observed a bimodal distribution, similar to the distribution reported for freely swimming {\it P.~putida} cells that exhibit both stops ($0^{\circ}$ turns) and reversals ($180^{\circ}$ turns).
For long events, the distribution is dominated by a peak at $180^{\circ}$, reflecting mechanical trapping, where a directional reversal is required to escape from the trap.
This finding supports our hypothesis that, besides mechanical trapping in the gel matrix, also actively triggered stop and turn events may occur as part of the swimming trajectories of {\it P.~putida} in semisolid agar.

\paragraph{Probability distribution of run- and dwell times.}

\begin{figure}[t]
    \includegraphics[width=\columnwidth]{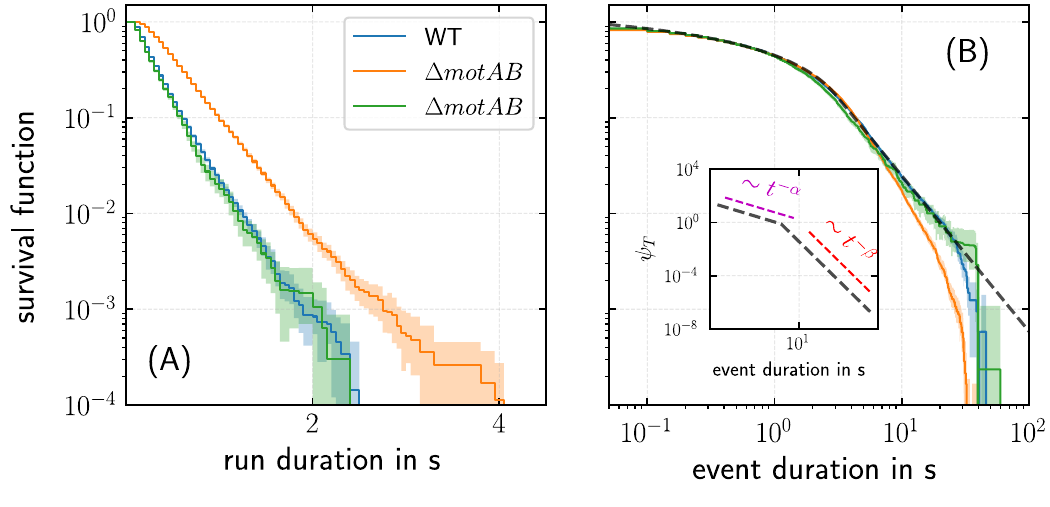}
    \caption{Survival functions~$\int_{t}^{\infty} dt' \, \psi_{R,T}(t')$ of the sojourn times of (A) runs and (B) events (the dotted line corresponds to a fit of the survival function of a piecewise power-law distribution~\cite{noteSI}); the inset in~(B) shows the piecewise power-law with the scaling of~$\psi_T(t)$ with~$\psi_T \propto t^{-\alpha}$ at intermediate timescales~($\alpha = 0.79$) and~$\psi_T \propto t^{-\beta}$ with $\beta=2.63$ for large times. Exponents were inferred by least-square fits to the survival function. We obtained the mean run times $\langle t_R \rangle = 0.35\, \mbox{s}, 0.52\, \mbox{s}\, \,\mbox{and}\, 0.32\, \mbox{s}$ and mean dwell times $\langle t_T\rangle = 2.07\, \mbox{s}, 1.77\, \mbox{s} \,\, \mbox{and}\, 1.94\, \mbox{s}$ for wild type, $\Delta$\textit{motAB} and $\Delta$\textit{motCD} cells, respectively, in $0.25\, \%$ agar. The shaded regions correspond to the error bars associated with the survival functions obtained by bootstrapping. }   
    \label{fig:duration}
\end{figure}

Based on the distinction between runs and events, we also determined the distribution of sojourn times in each state by non-parametric maximum-likelihood estimation~\cite{vardi1982nonparametric}. 
In the following, we denote these waiting time distributions~$\psi_R(t)$ and~$\psi_T(t)$, respectively. 
In Fig.~\ref{fig:duration}A, the distribution of run times is displayed in the form of a survival function. 
The semi-logarithmic plot reveals an exponential scaling, reflecting the random nature of the surrounding gel environment. 
The distribution of dwell times in immobile phases, in contrast, is well approximated by a piecewise power-law with an exponent~$\alpha < 1$ at intermediate timescales, followed by a steep cutoff towards larger dwell times with an exponent of $\beta > 2$, see Fig.~\ref{fig:duration}B and SM~\cite{noteSI}.

\paragraph{An active particle model for intermittent run motility.}

Our experimental observations suggest that \textit{P.~putida} cells in a gel-like environment resemble the motility of intermittently self-propelled particles. 
Based on a recently proposed modeling framework~\cite{datta_random_2024}, we describe the bacterial dynamics by two coupled Langevin equations
\begin{equation}
    \dot{\vec{r}}(t) = v \! \left (t \right ) \! \left(\begin{array}{c}\cos\phi(t) \\ \sin\phi(t)\end{array}\right) \! , \quad 
    \dot{\phi}(t) = \sqrt{2D \! \left(t\right)}\,\eta \! \left(t\right) \!,
    \label{eqn:model}
\end{equation} 
where the speed $v\! \left(t\right)$ and the rotational diffusion coefficient $D \! \left(t\right)$ of the bacterium are $\left(v_{0},D_{\phi}\right)$ in the run-state and $\left(0,0\right)$ during trap and turn phases; note that the spatial displacement during turns and traps is negligible compared to the displacement during runs. 
The duration of run and turn phases are drawn from the waiting-time distributions~$\psi_{R}(t)$ and~$\psi_{T}(t)$, respectively. 
Once the bacterium starts a new run, its direction of motion changes by a random angle~$\Delta \phi$ as determined experimentally~(Fig.~\ref{fig:angular_change}A). 
During runs, the direction of motion fluctuates, described by Gaussian white noise~$\eta(t)$ in Eq.~\eqref{eqn:model}. 
We denote the starting point of our observations as $t_a$, the so-called aging time.

For the model outlined above, we derived an analytic expression for the velocity auto-correlation function $C_{vv}\left(t_a;\Delta\right) = \left<\dot{\vec{r}}\left(t_a\right)\cdot\dot{\vec{r}}\left(t_a + \Delta\right)\right>$ in Laplace domain~\cite{datta_random_2024},
\begin{align}
    \begin{split}
    \hat{\hat{C}}_{vv} \! \left(s;u\right) &= \frac{v_0^{2}}{\left(u+D_{\phi}\right)} \left[\frac{1-\hat{\psi}_{R} \! \left(s\right)}{s \! \left\{1-\hat{\psi}_{R} \! \left(s\right)\hat{\psi}_{T} \! \left(s\right)\right\}} \right. \\
    &\left. - \frac{\hat{\psi}_{R} \! \left(u+D_{\phi}\right)-\hat{\psi}_{R} \! \left(s\right)}{\left\{s - \left(u + D_{\phi}\right)\right\}\left\{1-\hat{\psi}_{R} \! \left(s\right)\hat{\psi}_{T} \! \left(s\right)\right\}}\cdot \hat{G} \! \left(u\right)\right] 
    \label{eqn:Cvv_su}
    \end{split}  \nonumber
    \\
    \hat{G}\left(u\right) &= \frac{1-\Gamma \hat{\psi}_{T} \! \left(u\right)}{1-\Gamma \hat{\psi}_{T} \! \left(u\right)\hat{\psi}_{R} \! \left(u+D_{\phi}\right)}\, , 
\end{align}
where $\Gamma = \mean{\cos{\Delta\phi}}$ arises due to reorientations of the direction of motion; $s$ and $u$ are the Laplace variables corresponding to the aging time~$t_a$ and the lag time~$\Delta$, respectively. 
From this correlation function, we can find an exact expression for the MSD $m_2\left(t_a;\Delta\right) = \langle\left|\vec{r}\!\left(t_a+\Delta\right)-\vec{r}\!\left(t_a\right)\right|^{2}\rangle$ in the Laplace domain via
\begin{align}
    \hat{\hat{m}}_{2} \!\left(s;u\right) = \frac{2}{u} \cdot \frac{\hat{\hat{C}}_{vv} \!\left(u;u\right) - \hat{\hat{C}}_{vv} \! \left(s;u\right)}{s-u} \, ,
    \label{eqn:msd}
\end{align}
from which $m_2\left(t_a;\Delta\right)$ can be derived by numerically performing an inverse Laplace transform.  
To explicitly compute the MSD following Eqs.~(\ref{eqn:Cvv_su}, \ref{eqn:msd}), the parameters~$v_0, D_\phi$, and $\Gamma = \mean{\cos \Delta \phi}$ as well as suitable choices for the waiting time distributions~$\psi_{R,T}(t)$ are required. 
All of these can be readily derived from experimental data, enabling us to test the consistency of the model by comparing the experimentally observed MSD with the model prediction~(for details of estimation of parameters from the data, see SM~\cite{noteSI}).

\begin{figure}[t]
         \includegraphics[width=\columnwidth]{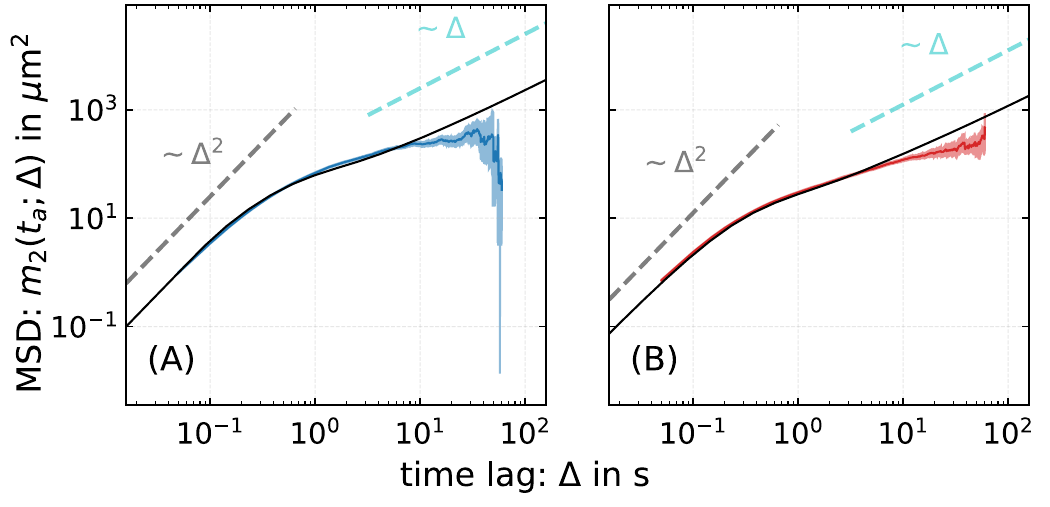}
         \caption{Comparison of MSDs predicted by the model~(black lines) with experimental data for \textit{P.~putida} wild type cells in (A)~$0.25\, \%$ agar~(blue) and (B)~$0.3\, \%$ agar~(red). The shaded regions correspond to the error bars ($95 \, \%$ confidence interval) associated with the MSD obtained by bootstrapping. } 
  	\label{fig:data_and_model}
 \end{figure}

In Fig.~\ref{fig:data_and_model}, we compare the experimentally measured MSD of {\it P.~putida} wild type cells in $0.25 \, \%$ and $0.3 \, \%$ agar and the predictions of the model, yielding quantitative agreement over more than two orders of magnitude in time. 
We note that the experimentally determined MSD curve for large lag times is increasingly unreliable due to the small number of long trajectories. 
In the long-time limit, the model predicts normal diffusion which is not resolved by the current experimental setup~\cite{Note1}. 
The model provides an analytical expression of the long-term diffusion coefficient for the bacteria undergoing intermittent run-motility 
\begin{align}
    D &= \frac{v_0^{2}}{2D_{\phi}} \frac{\mean{t_{R}}}{\mean{t_{R}}+\mean{t_{T}}}\left[1 - \frac{1-\hat{\psi}_{R} \! \left(D_{\phi}\right)}{\mean{t_R} \!D_{\phi}} \frac{1-\Gamma}{1-\Gamma \hat{\psi}_R \! \left(D_\phi\right)}\right] \! ,
    \label{eqn:diff_coeff}
\end{align}
where $\mean{t_R}$ and $\mean{t_T}$ are the mean values of the corresponding waiting time distributions~(cf.~caption of Fig.~\ref{fig:duration}). We predict a long-term diffusion coefficients~$D \approx 5.25 \, \mbox{\textmu m}^2 / \mbox{s}$ for wild type in $0.25\, \%$ agar, $2.56 \, \mbox{\textmu m}^2 / \mbox{s}$ for wild type in $0.3 \, \%$ agar as well as $8.55 \, \mbox{\textmu m}^2 / \mbox{s}$ for $\Delta$\textit{motAB} and $6.81 \, \mbox{\textmu m}^2 / \mbox{s}$ for $\Delta$\textit{motCD}, both in $0.25 \, \%$ agar.

\paragraph{Discussion.}

Single cell tracking of bacterial motion in complex environments has been a rapidly growing topic over the past years.
Examples include the swimming of bacteria, such as {\it E.~coli} or {\it Vibrio fischeri}, in two-dimensional chambers of different height~\cite{biondi1998random,lynch2022transitioning}, 
in narrow channels~\cite{vizsnyiczai2020transition,raza2023anomalous}, microfluidic mazes~\cite{weber2019rectification}, and in suspensions of micro-spheres~\cite{shrestha2023bacterial}.
Swimming bacteria have also been combined with lyotropic liquid crystals to form a novel type of living liquid crystal~\cite{zhou_living_2014,mushenheim2015effects}.
For the soil bacterium {\it P.~putida} considered in this work, swimming between planar interfaces~\cite{theves2015random} and in microfluidic chambers with cylindrical obstacles has been previously investigated~\cite{raatz2015swimming}.
Besides well-defined microfluidic geometries, bacterial swimming has also been studied in disordered porous and gel-like substrates.
In a porous matrix, formed by jammed packings of soft particles, {\it E.~coli} moved in directed paths that were interrupted by transient mechanical trapping events in narrow spacings or cavities~\cite{bhattacharjee2019bacterial,bhattacharjee2019confinement}, affecting also the chemotactic navigation strategy within the porous material~\cite{bhattacharjee2021chemotactic}.
Chemotaxis of the marine pathogen {\it Vibrio alginolyticus} was analyzed in agar hydrogels, showing that lateral flagella may prevent mechanical trapping and thereby improve the chemotactic performance~\cite{grognot_physiological_2023}.
Along with these experimental studies, also models were proposed to describe bacterial motility in heterogeneous disordered environments, see for example Refs.~\cite{licata2016diffusion,perez2021impact,lohrmann2023optimal,saintillan2023dispersion,kurzthaler2021geometric,datta_random_2024}.

Our present study of {\it P.~putida} motility in agar hydrogels showed that cells display intermittent run motility. 
As shown previously, the average swimming speed is only moderately decreased~\cite{Pfeifer2022} in comparison to locomotion in bulk liquid.
The duration of turn and trap events, during which bacteria are effectively immobile, are, however, increased by a factor of 10 on average compared to swimming in bulk liquid and show a power-law distribution with a steep cutoff towards large times.
Similar power-law scalings have been observed for {\it E.~coli} swimming in a porous material~\cite{bhattacharjee2019bacterial,bhattacharjee2019confinement}.
However, in contrast to the micron-sized openings between the surfaces of jammed particles that were used in these previous studies, in our case cells propagate though the disordered meshwork of an agar hydrogel. 
The average run times in our data are 10 times shorter compared to bulk swimming and exhibit an exponential distribution, see Fig.~\ref{fig:duration}. 
Compared to the wild type and the $\Delta$\textit{motCD} mutant, the run times of the $\Delta$\textit{motAB} mutant are longer on average.
Note, however, that the average swimming speed of the $\Delta$\textit{motAB} mutant is smaller than the speed of the wild type and the $\Delta$\textit{motCD} strain, see Supplemental Material~\cite{noteSI} and Ref.~\cite{Pfeifer2022}, so that traveling similar distances will require longer run times. 
Indeed, the run length distributions are identical for all three strains (see Supplemental Material~\cite{noteSI}), indicating that the distances traveled in the gel are mainly determined by the surrounding matrix.

These observations suggest that the swimming pattern in agar is affected by additional random trapping events in the polymer matrix.
An analysis of the turn angle distribution for events with short and long dwell times supports this hypothesis, see Fig.~\ref{fig:angular_change}. 
Together, these features of the swimming pattern result in a time evolution of the MSD that shows, after an initial ballistic regime, a~(transient) sublinear scaling, see Fig.~\ref{fig:trajectory}(B). 
This is reminiscent of subdiffusive behavior typically observed for transport in disordered environments~\cite{bouchaud1990anomalous,saxton2007biological,vilk2022unraveling}.
Examples include a wide range of different systems, such as passive particles diffusing in mucus or cytoskeletal networks~\cite{ernst2017model,wong2004anomalous}, water molecules at membranes~\cite{yamamoto2014origin}, active colloids in crowded environments~\cite{morin2017diffusion}, or transport in a lipid bilayer~\cite{akimoto2011non}.

To account for the experimentally observed scalings in the MSD, we rely on a recently proposed active particle model that is based on renewal theory, in which the active agents intermittently switch between two states:~a run state and immobile phases, during which active motion is absent but particle reorient their direction of motion~\cite{datta_random_2024}.
We performed extensive data analysis to extract the distributions of the run and turn duration as well as all other required parameters, such as the run speed, rotational diffusion coefficient and turn angles from data.
When supplied with the experimentally derived parameters and distributions, our model closely matches the experimental findings.
In particular, as a consequence of the power-law scaling in the waiting-time distribution~$\psi_T(t)$, the model recovers the transition from ballistic to a sublinear crossover regime observed in our data.
However, due to the cutoff of the power-law at large times, this regime remains transient.
In the long time limit, normal diffusion is predicted~\footnote{Note that the diffusive regime at long times is not observed in our data due to experimental limitations. With increasing time, it becomes more likely that typical bacterial swimming trajectories will leave the field of view accessible by a conventional microscopy setup. For this reason, long-time recordings selectively focus on trajectories of trapped or less motile cells that remain in the field of view, thus concealing the crossover to normal diffusion.}.

Dynamical switching between motile and resting phases has been reported for a range of different systems, such as active colloidal particles in the presence of AC fields~\cite{karani2019tuning,pradillo2019quincke}, stick-and-slip dynamics of myxobacteria~\cite{gibiansky2013earthquake}, near-surface swimming of {\it E.~coli}~\cite{perez2019bacteria}, or navigation strategies of larger organisms~\cite{nathan2008movement}.
Also on a conceptual level, similar dynamics has been considered for diffusive particles that switch between mobile and immobile states~\cite{doerries2023emergent} or for active Brownian particles alternating between stop and go phases~\cite{peruani2023active}.
Here, we have shown that bacterial swimmers in a gel matrix follow a similar intermittent run dynamics that can be described by a renewal process with power-law distributed dwell times.
Given that many common bacterial habitats exhibit gel-like properties, we expect these findings to advance our quantitative understanding of how bacterial swimmers invade complex environments such as the human gut or the plant rhizosphere. 
How this movement strategy impacts chemotactic navigation in heterogeneous environments remains an open question to be addressed in future work.

\begin{acknowledgments}

\paragraph{Acknowledgments.}

This research has been partially funded by Deutsche Forschungsgemeinschaft~(DFG): Project-ID 318763901~--~SFB1294. 

\paragraph{Author contribution.}

SB and VP conducted the experiments. AD, RG and SB processed the experimental data and performed the analysis. AD and RG did the modeling. CB, AD and SB drafted the manuscript. VP wrote the experimental methods. RG and CB supervised the whole project. All authors discussed the results and contributed to the final manuscript.

\end{acknowledgments}

%

\end{document}